\documentclass[12pt,a4paper]{book}
\usepackage[a4paper,margin=2cm]{geometry}
\pdfoutput=1
\usepackage{setspace}
\usepackage{verbatim}
\usepackage[ansinew]{inputenc}
\usepackage[french]{babel}
\usepackage[cam,center]{crop}
\usepackage{eurosym}
\usepackage{color}
\usepackage{microtype}
\usepackage{amsmath}
\usepackage{fancyhdr}
\usepackage{longtable}
\usepackage[final]{pdfpages}
\usepackage{fancybox}

\usepackage[OT2,OT1]{fontenc} \newcommand\cyr
{
\renewcommand\rmdefault{wncyr} \renewcommand\sfdefault{wncyss} \renewcommand\encodingdefault{OT2} \normalfont
\selectfont
}
\DeclareTextFontCommand{\textcyr}{\cyr}    
\usepackage{afterpage}
\usepackage[strict]{changepage}

\begin{document}

\thispagestyle{empty}

\begin{center}
\begin{Large}
{\textbf{UNIVERSITEIT UTRECHT}}
\end{Large}
\mbox{   }\\
\mbox{  }
\\ \mbox{} \\
\textbf{\large SCRIPTIE}\\
DISSERTATION
\\ \mbox{} \\
in Geophysics
\\ \mbox{} \\
{\large  Discrimination between quarry blasts and
micro-earthquakes using spectral analysis, applied to local Israeli events}
\\ \mbox{} \\
by :
\\ \mbox{} \\
Michiel Arjen Benjamin POSTEMA 
\\ \mbox{} \\
supervisor : Dr Torild van Eck 
\\ \mbox{} \\
submitted : July 16, 1996
\\ \mbox{} \\
\mbox{}
\end{center}
\newpage

\chapter*{1. Introduction}
During the last ten years a growing interest has been shown in discriminating man-made seismic activities from small earthquakes at regional distances. This interest is mainly associated with the negotiations on a Comprehensive Test Ban Treaty (CTBT) (Stump, 1991). Although events with
magnitudes larger than about 3 are said to be identified and verified easily, for smaller events no conclusive identification method has yet been found. Usually the problem is split into four parts: detection, identification, verification and discrimination. Detection is defined as recognizing signal from an event as `not being noise'. Identification is the process of finding out whether a seismic signal is caused by an earthquake or a certain man-made explosion, by means of positive deduction. During the verification concrete proof for the identification is to be found. Discrimination can be
seen as an identification method by negative deduction. If for example an event can certainly not be identified as an earthquake or a quarry blast, this event might be a nuclear test.

The new test ban treaty aims at banning all nuclear explosions down to 1~kt. Future verification efforts within the test ban treaty context will focus on the Middle East, among others Israel (Van
Eck, 1995).

One of the ideas for discriminating seismic events is to identify spectral modulation resulting
from time-delay shooting (ripple-firing). Earlier studies (Gitterman \& Van Eck, 1993; Hedlin et al., 1989; W\"{u}ster, 1996) had to make use of high-frequency data to identify this spectral modulation.

This study presents the concept of spectral modulation and a time-frequency analysis, applied to broadband local ($5<\Delta<200$\,km) seismic data from quarry blasts and micro-earthquakes, kindly supplied by the Institute for Petroleum Research and Geophysics (IPRG), Holon, Israel, from the so-called GIF-array. The aim of this research is verification of ripple-firing by recognition of scalloping trends in amplitude-spectra and hence discriminating quarry blasts from other events. Most quarry blasts are ripple-fired, in northern Israel open pit blasts; consequently the event discrimination method based on the recognition of ripple-firing patterns in the signal was chosen.

The methods presented in this paper might also be applicable for deconvolution and dereverberation purposes in exploration seismology.

\begin{figure}
\begin{center}
\includegraphics[width=0.3\textwidth]{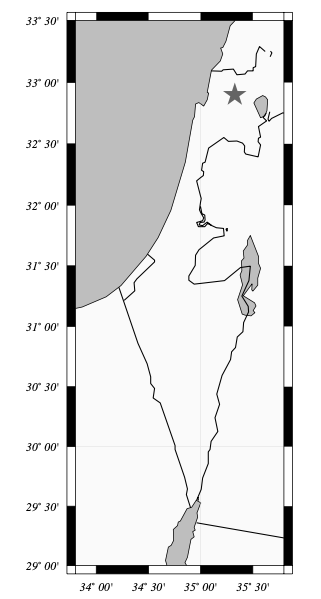}
\caption{Location of the GIF-array}
\end{center}
\end{figure}

\newpage

\chapter*{2. Theory}
In this section the characteristics of spectra of ripple-firing events are derived. Furthermore the theory behind automatic detection of reverberations is dealt with.

\section*{2.1 Frequency analysis}

If the S-N ratio is high, a reverberated analogue seismic time-domain signal $s(t)$, in which the reverberation is caused by the source, can be expressed by the following equation (cf. figure~2):
\begin{equation}
s(t) = \mathcal{U}(t) * e(t) + \varepsilon(t) = u(t) * h(t) * e(t) + \varepsilon(t) \approx u(t) * h(t) *e(t)
\end{equation}
\begin{align*}
\textrm{in which:} ~~~& \mathcal{U}(t)& \textrm{source function} \\
				& u(t)		& \textrm{source function for a `single' source} \\
				& h(t)		& \textrm{reverberation function}\\
				& e(t)		& \textrm{earth and instrument response}\\
				& \varepsilon(t)	& \textrm{additive noise}
\end{align*}

\begin{figure}[h!!!!]
\includegraphics[width=\textwidth]{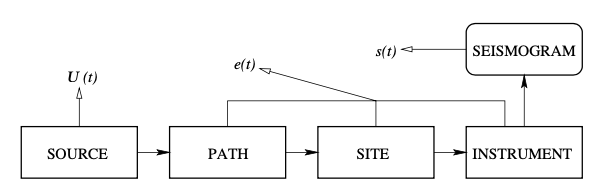}
\caption{Schematic representation of the evolution of a seismic signal}
\end{figure}

Figure 3 shows an example of ripple-firing geometry in an open pit (Smith, 1993). Triangles indicate explosives. The individual rows of explosives are detonated with a certain time-delay, to
enlarge the impact of the event.
In such a case the source function is expressed as follows:
\begin{equation}
U(t) = u(t) + \sum_{n=1}^N (1-a_n) u(t-n\Delta t+\theta_n) = u(t) * \left(   \sum_{n=1}^N (1-a_n) \delta(t-n\Delta t+\theta_n)   \right)
\end{equation}
\begin{align*}
\textrm{in which:} ~~~& \delta(t'-t)	& \textrm{Dirac's delta-function} \\
				& a_n		& \textrm{relative yield variation per row}~n \\
				& \Delta t		& \textrm{seismic travel-time between two ripples}\\
				& N			& \textrm{total number of ripples}\\
				& \theta_n		& \textrm{perturbation in}~ \Delta t
\end{align*}
and in practise:
\begin{equation}
\Delta t = \tau - \frac{l}{v} \cos \alpha \approx \tau
\end{equation}
\begin{align*}
\textrm{in which:} ~~~& \tau		& \textrm{time-delay between two row-detonations} \\
				& l			& \textrm{distance between shot-arrays} \\
				& v			& \textrm{wave-velocity}\\
				& \alpha		& \textrm{angle of shotpoint configuration}
\end{align*}
The traveltime-delay of the individual signals caused by explosions within a row is negligible in
relation to the time-delay between two row-detonations. If the relative yield variation per row and
the perturbations in $\Delta t$ are small the reverberation function is approximated by a finite Dirac's comb:
\begin{equation}
h(t) = \sum_{n=0}^N (1-a_n) \delta(t-n\Delta t + \theta_n ) \approx \sum_{n=0}^N \delta(t-n\Delta t)
\end{equation}

\begin{figure}
\begin{center}
\includegraphics[width=0.7\textwidth]{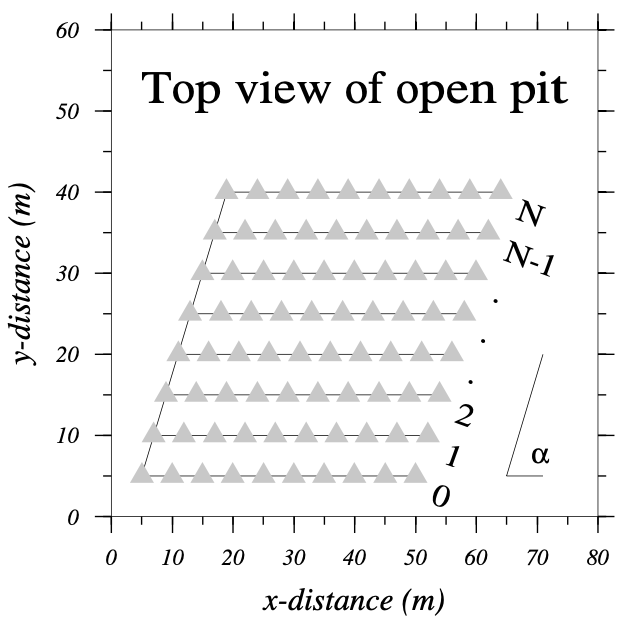}
\caption{Example of a simple ripple-firing geometry in an open pit}
\end{center}
\end{figure}

The approximations made are necessary for deriving the characteristics of ripple-firing signals in the frequency-domain. In common cases the time-delay between two explosions is: $10 \le \tau \le 70$\,ms. In the frequency-domain the total signal of equation~1 becomes:
\begin{equation}
S(f) = U(f) H(f) E(f)
\end{equation}

The derivation of the reverberation spectrum is analogue to the derivation of the Dirichlet spectrum (Van den Berg \& Neele, 1993).
\begin{align}
H(f) & = \int_{-\infty}^{\infty} \sum_{n=0}^N \delta (t-n\Delta t) e^{2\pi ift} dt = \sum_{n=0}^{N-1} e^{2\pi inf\Delta t} = \frac{1 - e^{2\pi iNf\Delta t} }{ 1 - e^{2\pi if\Delta t} } = \\
& = \frac{e^{\pi i Nf\Delta t}}{e^{\pi i f\Delta t}} \frac{ e^{-\pi i Nf\Delta t} - e^{\pi i Nf\Delta t} }{ e^{-\pi i f\Delta t} - e^{\pi i f\Delta t} } = e^{\pi i (N-1) f\Delta t} \frac{ \sin(\pi Nf\Delta t) }{ \sin(\pi f \Delta t) } \nonumber
\end{align}
where $i$ is the complex number with the property $i^2 = -1$.

For the amplitude spectrum this implies:
\begin{equation}
|H(f)| = \left|   \frac{ \sin(\pi Nf\Delta t) }{ \sin(\pi f \Delta t) }   \right|
\end{equation}

\begin{figure}
\begin{center}
\includegraphics[width=0.9\textwidth]{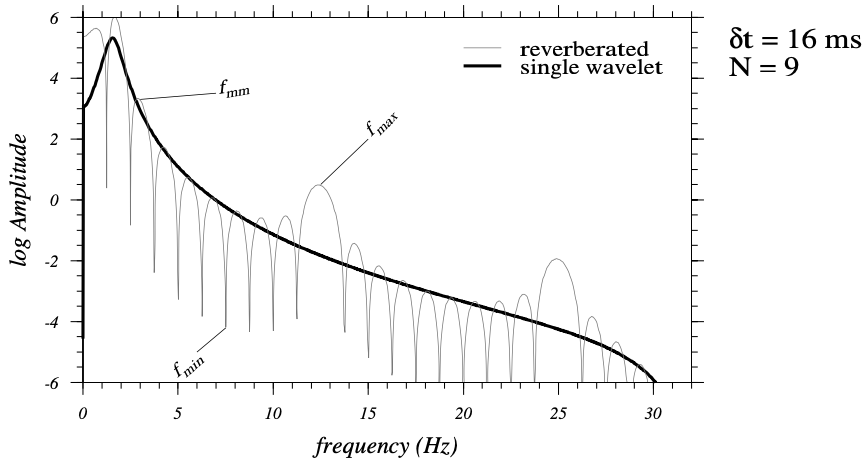}
\caption{Amplitude-spectrum of a synthetic ripple-firing blast with $\Delta t = 16$ ms and $N=9$}
\end{center}
\end{figure}

A synthetic function $\log |U(f)||H(f)|$ is graphically shown in figure 4. For $u(t)$ a single minimum-phase wavelet was defined by the Berlage function (B{\aa}th, 1974) 
$B(t) = \mathcal{H} (t)   t  e^{-at} \sin \omega_0 t $, 
with $\mathcal{H}(t)$ being the Heaviside function and $a$ and $\omega_0$ being positive constants. 
The obviously visual trend is called scalloping. Its characteristics can easily be derived (Gitterman \& Van Eck, 1993):
\begin{align}
f_{max} = \frac{k}{\Delta t} 				& ~~~~~~~~~~& k~ \in~ \{0,1,2,...\}\nonumber\\
f_{mm} = \frac{ k + \frac{1}{2} }{N \Delta t }  	& & k ~\in ~\{1,2,...,N-2,N+1,N+2,...\}	\\
f_{min} = \frac{ k }{ N\Delta t } 				& & k ~\in ~\{1,2,...,N-2,N+1,N+2,...\} \nonumber
\end{align}
The values of $f_{max}$ are defined as the frequencies at which both the functions $\sin (\pi N f \Delta t)$ and $\sin (\pi  f \Delta t)$ are equal to zero. The values of $f_{mm}$ represent those frequencies at which the function $|\sin (\pi N f \Delta t)|$ has a maximum. The frequencies at which the notches occur are those where $\sin (\pi N f \Delta t)=0$ \emph{and} $\sin (\pi  f \Delta t) \neq 0$.

Thus, if the Fourier transform $S(f)$ of a seismic signal contains a certain scalloping trend which proves not to be caused by path- or site-effects, conclusions can be drawn about reverberation in
the source.

It should be mentioned that a silent assumption has been made. The individual waves were
considered as response functions, forming a black box (cf. figure 5). The response functions in the
black box are assumed to be continuous and varying slowly in the frequency domain, compared with scalloping. This way scalloping should still be detectable among the trends in the amplitude-spectra.

\begin{figure}
\begin{center}
\includegraphics[width=0.9\textwidth]{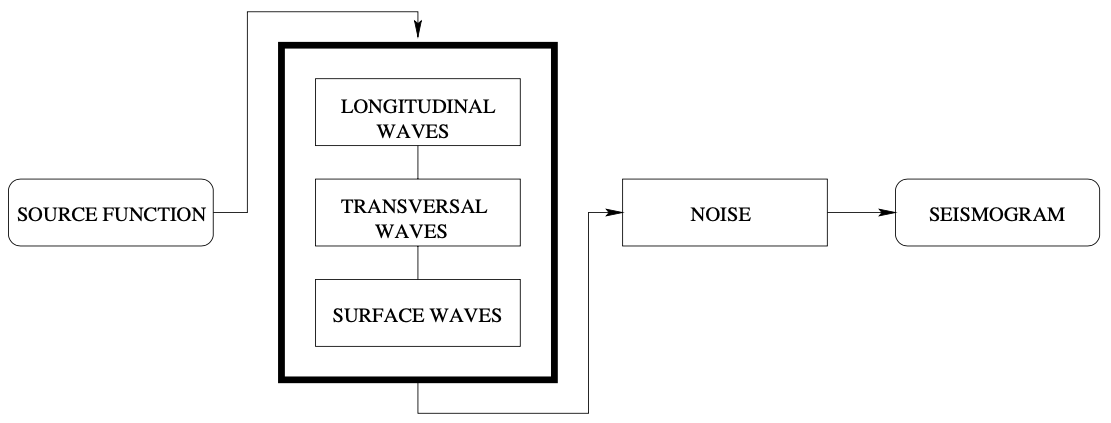}
\caption{Schematic representation of the ``black box'' system}
\end{center}
\end{figure}

\section*{2.2 Methods to recognize scalloping}

\subsection*{2.2.1 Complex cepstral analysis}

Taking the logarithm of the frequency-spectrum will cause the multiplied functions in equation~5
to be separated into individually added components:
\begin{align}
\hat{S}(f) ~ & \equiv \log S(f) = \log [ U(f) H(f) E(f) ] = \log U(f) + \log H(f) + \log E(f) \equiv \\
		&   \equiv \hat{U}(f) + \hat{H}(f) + \hat{E}(f)   \nonumber
\end{align}
As was shown earlier $H(f)$, and hence $\hat{H}(f)$, is periodic with $\frac{1}{\Delta t}$ and  $\frac{1}{N\Delta t}$.
Therefore the inverse Fourier transform of  $\hat{H}(f)$ is expected to be nonzero only at integer multiples of $\Delta t$; $N\Delta t$  already is an integer multiple of $\Delta t$.

The inverse Fourier transform of the logarithm of a frequency-spectrum is called the complex cepstrum. It operates in the quefrency-domainwith unit Zhert (equivalent to $\frac{1}{Hz}$=sec), fully similarly to the time-domain.

Applying this theorem to the case above leads to:
\begin{equation}
\hat{h} (\xi) = \frac{1}{2\pi} \int_{-\infty}^\infty \hat{H}(f) e^{-2\pi if\xi} df = \sum_{-\infty}^{\infty} c_n \delta(\xi - n\Delta t)
\end{equation}
\begin{align*}
\textrm{in which:} ~~~& \xi			& \textrm{quefrency} \\
				& c_n		& \textrm{nonzero constants} \\
				& \Delta t		& \textrm{expressed in unit Zhert}
\end{align*}
Hence a means seems to be provided to detect the properties of reverberation.

However, if the (complex) function $H(f)$ is written as a combination of its length and its
argument, another problem appears.
\begin{equation}
H(f) = |H(f)| e^{i \arg H(f)} = |H(f)| e^{i \, \mathtt{ARG} \, H(f)}
\end{equation}
and on the other hand
\begin{equation}
\hat{H}(f) = \log |H(f)| + {i \arg H(f)} \neq \log  |H(f)| + {i \, \mathtt{ARG} \, H(f)}
\end{equation}
$\mathtt{ARG} \, H(f)$  is the apparent argument with range $]-\pi,\pi]$.
The function $\mathtt{ARG} \, H(f)$ is discontinuous, thus not analytic. 
Therefore it cannot satisfy the requirements for the inverse transform, unless a so-called phase-unwrapping procedure (Oppenheim \& Schafer, 1975) is applied to 
$\mathtt{ARG} \, H(f)$, with which the true $\arg H(f)$ is to be retrieved. 
This is possible if and only if the phase varies slowly compared with the intermediate distance of the frequency-samples.

\subsection*{Cross-correlation of data from different events}

The location of an arbitrary event should be verified simply by comparing the seismic signal with those of different events, recorded by the same station. Similarities of time-domain signals can be detected both in the time-domain and in the frequency-domain by writing the cross-correlation function
$\theta_{12}$ as a convolution:
\begin{equation}
h(t) = f(t) * g(t) = \int_{-\infty}^{\infty} f(\tau) g(\tau-t) d\tau \longleftrightarrow H(f) = F(f) G(f)
\end{equation}
\begin{equation}
\theta_{12} (t) = \int_{-\infty}^{\infty} f(\tau) g(\tau-t) d\tau = \int_{-\infty}^{\infty} f(\tau) g(-(\tau-t)) d\tau = f(t) * g(-t)
\end{equation}
Because the time-signals are real, it follows that $|G(-f)| = |G(f)|$ and hence:
\begin{equation}
\Theta_{12} (f) = F(f)  G(-f) = |F(f)|  |G(f)|  e^{i[ \arg F(f) - \arg G(f) ]} = F(f) G^* (f)
\end{equation}
where $G^* (f)$ is the complex conjugate of $G(f)$

It should be verified that eventual scalloping features are not caused by path- or site-effects. One
method of recognizing the contribution of the latter is applying a cross-correlation on amplitude-spectra of data of different quarry blasts which occured in the same pit, recorded by the same
station. Similarly to the above case the cross-correlation can be performed both in the frequency-domain and in the time-domain:
\begin{equation}
\theta_{12} (f)   =  F(f) * G(-f)   \longleftrightarrow \Theta_{12} (t) = f(t) g^* (t)
\end{equation}

\newpage
\chapter*{3 Data and methods}

\begin{figure}
\begin{center}
\includegraphics[width=0.5\textwidth]{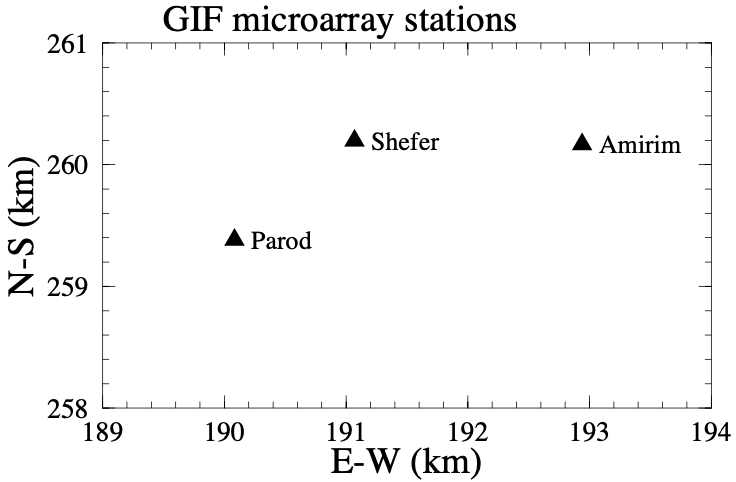}
\caption{Configuration of the GIF-array}
\end{center}
\end{figure}

The seismograms of 90 events in northern Israel, which occurred between 23 October and 21 December 1994, were available for the analyses described in the last section. The data were
recorded by a 3-component, 3-station array. The locations of the stations (expressed in ICCS) are respectively: Shefer -- 191.069E, 260.879N; Parod -- 190.83E, 259.382N; Amirim -- 192.939E, 260.197N (cf. figure 6).

Table 4 shows a list of the data used, including the (probable) identifications, locations and distances to the array. In the identification column ``XY'' means ``confirmed explosion''.

The open pits present in the neighbourhood are Amiad (201E, 258N), Arabel (181E, 240N), Golani (188E, 244N), Kadarim (193E, 258N) and Yehiam (172E, 271N).

The sample-rate of all traces was 16~ms. For this Levant area the broadband data, containing frequencies up to 31.25~Hz, could be considered unique. The S--N ratio was often less than~0.7, leaving the P- and S-arrivals hardly visible. It was decided that only the seismograms having a relatively large S--N ratio and being recorded by all of the three stations would be used for the analyses.

Frequency-spectra were obtained by performing Fast Fourier Transforms on the data. Up to 20~Hz the spectra of the signals used, differ significantly from the noise-spectra. Figure~12 shows examples of an open pit blast and an earthquake respectively.

\begin{figure}
\begin{center}
\includegraphics[width=0.9\textwidth]{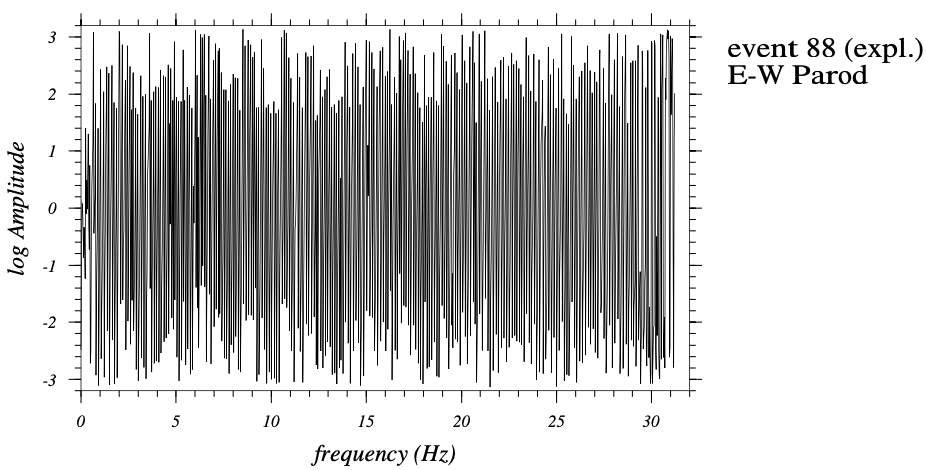}
\caption{Example of a phase-spectrum of an explosion}
\end{center}
\end{figure}

Although the signals contain high frequencies unique for the Levant, they are still not comparable with the very high-frequent data recorded in some other areas (W\"{u}ster, 1996), even reaching 500~Hz! The phase-spectra are varying very fast compared with the intermediate distance of the frequency-samples, which makes phase-unwrapping, and hence complex cepstral analysis, very difficult (cf. figure~7).

Figure~13 illustrates the flow-chart of the analysis system used in this research. The frequency analysis consists of performing FFT, determining the S--N ratio and possibly applying smoothing. The minimal difference between signal- and noise-amplitude, the window-length used and a smoothing factor may be used as indicators. The analysis can be performed in 3D by making a spectrogram or surface plot.

\begin{figure}
\begin{center}
\includegraphics[width=0.9\textwidth]{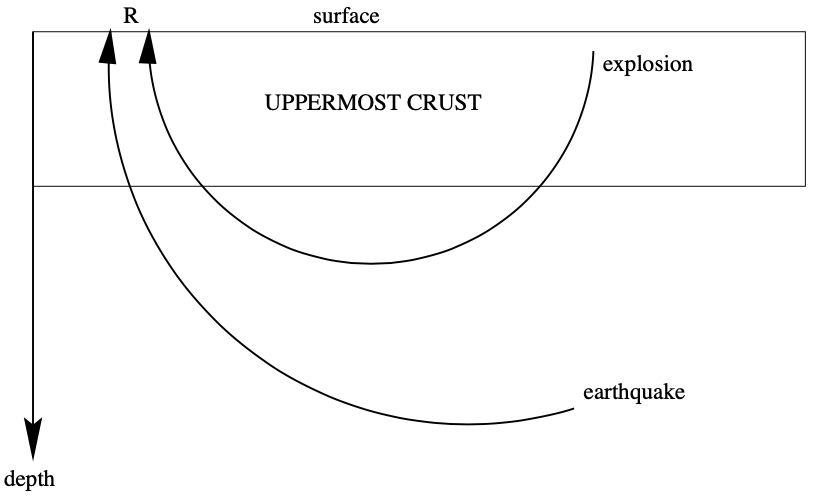}
\caption{Schematic cross-section}
\end{center}
\end{figure}

Explosions are expected to have a lower frequency-content than earthquakes. The reason for this is clarified in figure~8. The uppermost crust contains less dense material than the crust underneath, which results in a high attenuation of seismic waves going through. Because explosions occur near the earth surface, their signals travel through this medium twice, earthquake signals, commonly generated much deeper in the seismogenic zone between~4 and 20~km, only once. Thus the attenuation, and hence the lower frequency-content, of explosions is larger. This way the probability (P) that a seismic signal is either an explosion or an earthquake can be estimated.

Open pit blasts happen only on a limited number of locations. The seismic signals generated
in one pit are assumed to be travelling over the same path and are therefore expected to show a
high time-domain correlation with each other.

Several events recorded had already been identified and located. With the data of these events
(the so-called data-bank) both the locations and the contribution of path- and site-effects several arbitrary events should be estimated by means of the methods described in \S 2.2.2.

Spectral smoothing was done by adding the Amplitude-spectra of the different stations for each component. Applying smoothing windows to the spectra would not have been useful, because the equidistant notches $f_{min}$ in the amplitude-spectra (cf. figure 4) should remain visible.

The advantages of adding components are the reduction of the variance and the improvement of the imaging of raw features in spectra.

\chapter*{4 Results and discussion}
In this chapter results are shown of the location estimations and of the scalloping verification effort. As stated above, homomorphic deconvolution by cepstral analysis was not succesfull due to the rapidly varying phase-spectra.

\section*{4.1 Time-domain cross-correlation}

Location verification was performed on some high S--N explosion events, all with suspected or known locations in the Galilee/Kinneret region, by cross-correlating the time-signals of those events with each other. The results of vertical component data from Parod are shown in table 1.
Events 3, 26 and 48 are confirmed Kadarim explosions. Event 10 was confirmed by Amiad. Events 21 and 41 are suspected to be Amiad events too, although 41 had been confirmed by Kadarim!
Finally events 36, 62 and 70 were identified as Galilee explosions, probably having occurred in Kadarim. The results of the cross-correlation with the Amiad data did not differ more than 0.04
in comparison with the values in table 1. The time-domain data recorded by Shefer were not considered because of the high noise-level.

\clearpage
\mbox{}\\Table 1: Maxima of time-domain cross-correlation

\includegraphics[width=0.8\textwidth]{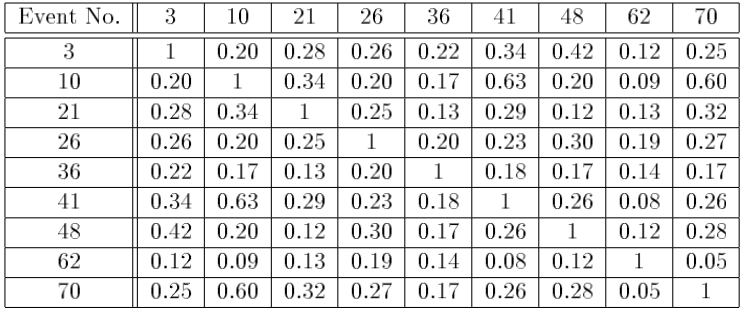}

Normally time-domain signals recorded by an array are stacked per component, but in this case it is not an alternative because of the low coherency between the stations, expressed in table 2.

\mbox{}\\Table 2: Cross-correlation between Parod and Amirim stations

\includegraphics[width=0.7\textwidth]{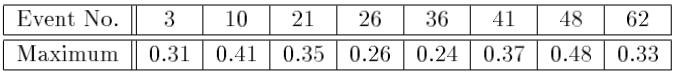}

The low correlation of events in table 1 may be explained by the low correlation between the stations for individual events. This leads to the conclusion that the influence of the source site-response, due to small variations of the source locations is high and is difficult to neglect during
the analyses.

\section*{4.2 Frequency-spectra and auto-correlation}

\begin{figure}
\begin{center}
\includegraphics[width=0.9\textwidth]{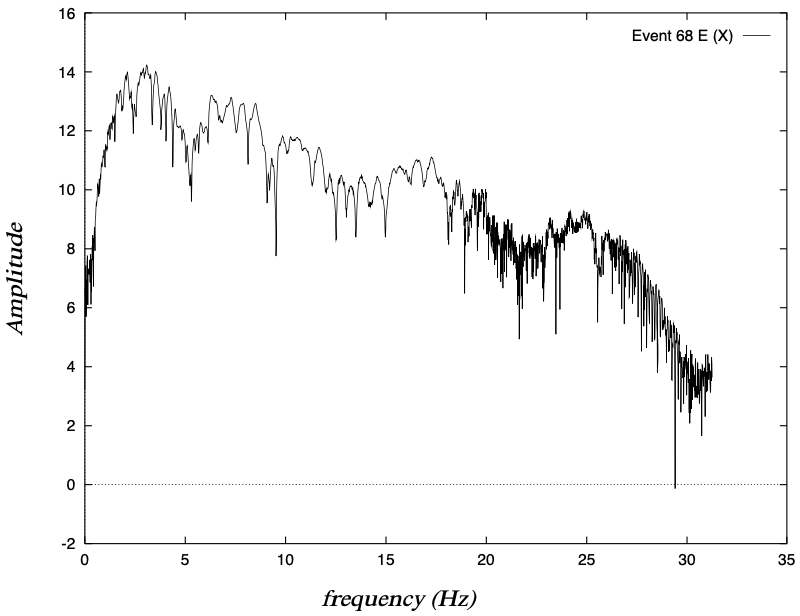}
\caption{Amplitude-spectrum with obvious scalloping trend}
\end{center}
\end{figure}

Some of the spectra of explosion signals seem to contain a visual kind of scalloping. Figure 9 shows an example of such a spectrum. Performing auto-correlations, either on individual components or on added components of amplitude-spectra, should provide an automatic means for detection of scalloping.

Log(amplitude-spectra) of 44 events were calculated after which the spectra were added per com- ponent. Thus for each event (3+1)3=12 spectra were available for autocorrelation analysis. Only the parts of the spectra between 1 Hz and 25 Hz were taken into consideration, because of the high noise levels and the deviating linear trends outside those limits.

Before the auto-correlations means and linear trends were removed from the log(spectra) to avoid `boxcar-effects' and disturbance from trends caused by wavelet- and path-functions.

\begin{figure}
\begin{center}
\includegraphics[width=1.0\textwidth]{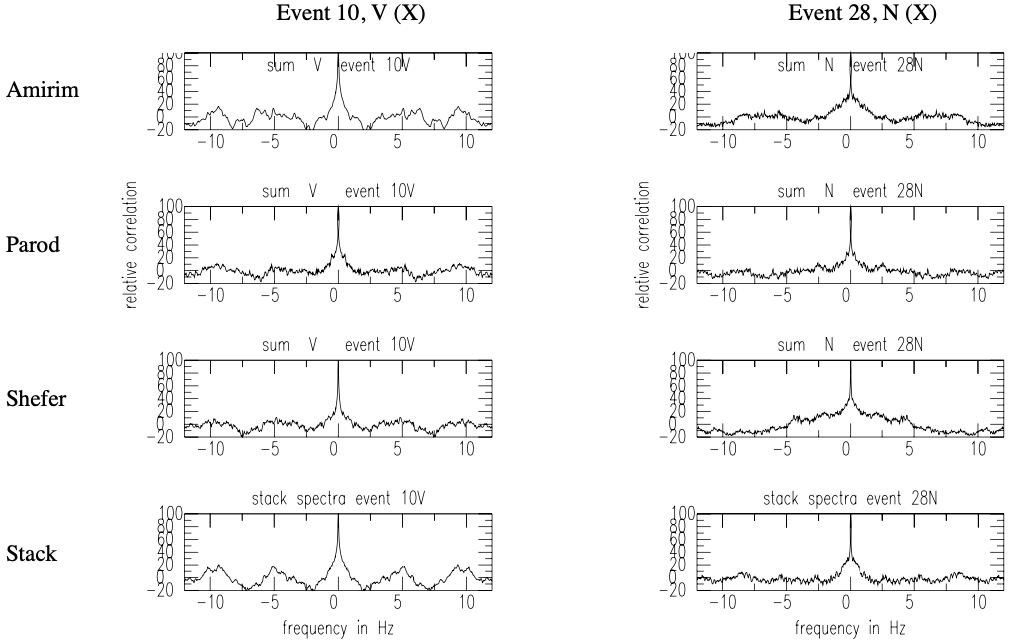}
\caption{Correlograms. Left \{+,+,+,+\}, right \{+,v,v,+\} }
\end{center}
\end{figure}

The combined auto-correlations of an event must provide a criterium for scalloping verification.
This criterium is concerned with the maximum height of a repetition peak. Initially it was assumed
that max(repetition peak) should be over 10\% of the zero-peak to confirm scalloping, but after
the first runs this was adapted to a 7$\frac{1}{2}$\% criterium. Figure 10 shows an example of a correlogram which obviously satisfies the criterium.
\clearpage
\mbox{}\\
Table 3: Results of auto-correlations

\includegraphics[width=0.7\textwidth]{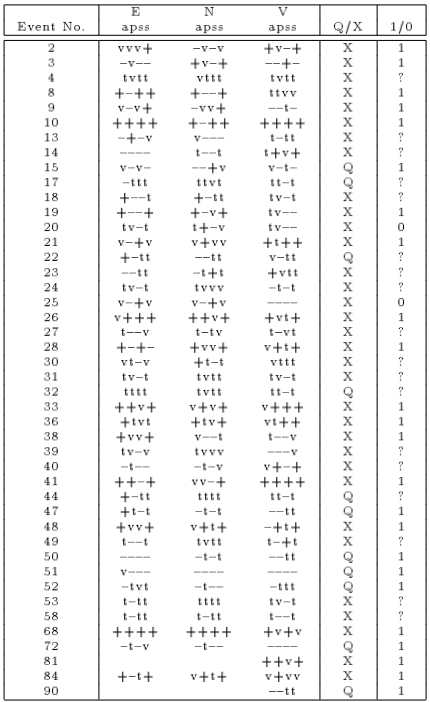}

For evaluation purposes of the correlograms four categories were defined: satisfying the criterium (+), containing visual repetition but not satisfying the criterium (v), not decisive due to trends in the log(amplitude-spectrum) (t), and not containing visual repetition (-). These grades were applied to the stations Amirim, Parod, Shefer and the stack (apss) for every east, north and vertical component for every event used. Examples of these categories are shown in figures 10 and 11. Table 3 illustrates these results. Q/X shows whether an event is an earthQuake or an eXplosion. The 1/0 column indicates the success of this experiment. 1 means that the nature of an event can be derived from the qualifications given (+,v,t,-), 0 means the opposite, and ? stands for indecisive.

\begin{figure}
\begin{center}
\includegraphics[width=1.0\textwidth]{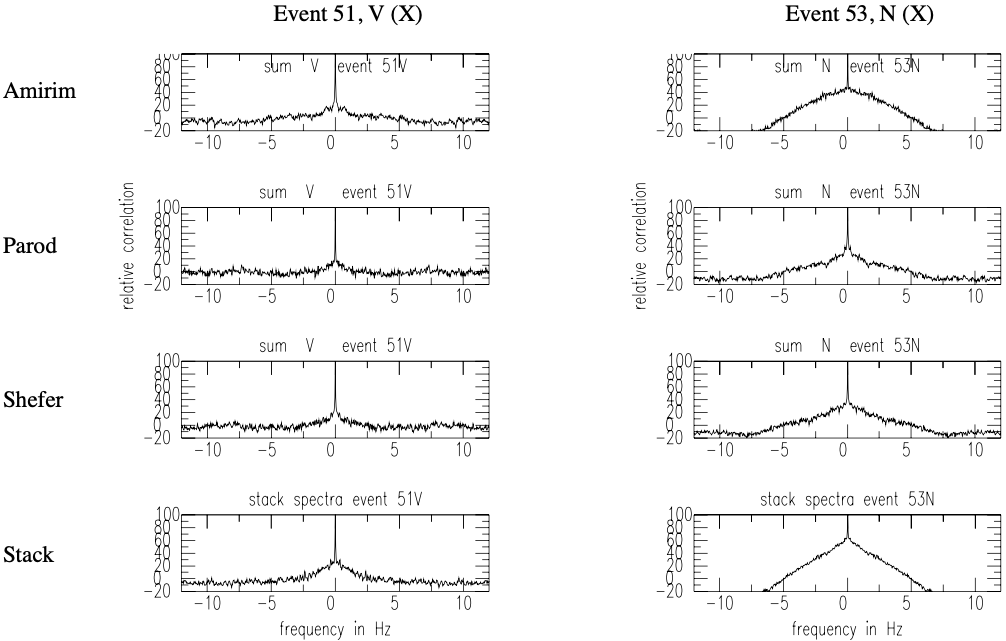}
\caption{Correlograms. Left \{-,-,-,-\}, right \{t,t,t,t\} }
\end{center}
\end{figure}

Evaluation of the correlograms and selection by weighting plusses and `v's against minus-signs
(`t's are neutral) were done manually. Both these difficult tasks are to be done automatically. The system used was as follows: Each symbol was given a mark (+=1, v=0.5, t=0, -=-1), the marks for stacked components were added. If the result of this sum equalled 1 or higher, the success-rate was set to 1, while it was set to 0 in case of -1 or lower. Everything in between, rated ``indecisive'', was submitted to a subjective second analysis, in which the general impression from the other components was used as judge.

From the 44 events, 24 can automatically be identified and verified with this method (1), 18 need further attention (?) and only 2 are invalid (0).

Although the results appear to be very positive, it should be taken into account with the evaluation of the method presented in this paper that still no ``formula'' has been provided for automatic verification. The borders between the categories are vague and partially overlapping. Moreover the grade ``visual'' is too subjective for automatic implementation. In other words: the results are not considered as being decisive.

\chapter*{Conclusions}
The analysis system demonstrated in figure 13 appears to be a good estimator for indentification and verification purposes. For high S--N events locations can hardly be confirmed with time-domain cross-correlation, because the influence of the source site-response is difficult to neglect.

Complex cepstral analysis was not performed due to the rapidly varying phase-spectra of the events used.

The method utilizing auto-correlation of amplitude-spectra presents suprisingly good results, although the definition of the categories and the weighting is a very subjective operation.

In many spectra scalloping features are visible, but still it is questionable whether an automatically applicable detector can and will be develloped.

\chapter*{6 Future issues}

In this research characteristics of noise were usually ignored. Auto-correlations should be per- formed on noise-spectra as well, and be compared with the results obtained from the signals, for some of the reverberations observed might be noise-induced.

For reasons concerned with tidyness conventional band pass filtering should be applied to the time-domain windows before correlation.

The trends in log(amplitude-spectra) which cause triangle-effects in correlograms can obviously not be corrected by linear trend removal only. In other words, deconvolution with the fnction log
$\frac{1}{f^n}$ is not sufficient. Other functions will have to be used instead.

Attention will have to be spent on the subject of decoupled explosions. It might be possible that seismic echoes produced by cavity walls cause the same kind of scalloping as the reverberations from ripple-firing explosions.

\chapter*{7 Acknowledgements}

This research was supervised by Dr. Torild van Eck, for which I am very grateful. Prof. Michael Hedlin sent me a computer program for complex cepstrum calculation and Dr. Jan W\"{u}ster supplied me with his up-to-date work on this subject. Many thanks to both! As mentioned before the data were kindly supplied by the IPRG. 

\chapter*{8 List of abbreviations}
\begin{tabular}{l|l}
CTBT&Comprehensive Test Ban Treaty\\
FFT & Fast Fourier Transform\\
GIF & German Israeli Foundation\\
GMT & Greenwich Mean Time\\
ICCS & Israel Cartesian Coordinate System \\
IPRG & Institute for Petroleum Research and Geophysics \\
S--N & Signal to Noise
\end{tabular}

\chapter*{References}
B{\aa}th, M.: Spectral Analysis in Geophysics, Chap. 2, p. 53, Elsevier Scientific Publishing Company, Amsterdam, The Netherlands, 1974.
\\

Berg, A.P. van den, F.P. Neele: Geofysische Dataverwerking, Utrecht, The Netherlands, 1993.
\\

Dobrin, M.B., C.H. Savit: Introduction to Geophysical Prospecting, 4th ed., Chap. 6, pp. 174--183, McGraw-Hill Book Company, Singapore, 1988.
\\

Eck, T. van: Nuclear test ban treaty verification: Relevant seismological research topics for Israel, Report to the IAEC, preliminary version (95AOl), Utrecht, The Netherlands, 1995.
\\

Gitterman, Y., T. van Eck: Spectra of quarry blasts and microearthquakes recorded at local distances in Israel, Bull. Seism. Soc. Am., 83, pp. 1799--1812, 1993.
\\

Hedlin, M.A.H., J.B. Minster, J.A. Orcutt: The time--frequency characteristics of quarry blasts and calibration explosions recorded in Kazakhstan, USSR, Geophys. J. Int., 99, pp. 109--121, 1989.
\\

Hedlin, M.A.H., J .B. Minster, J .A. Orcutt: An automatic means to discriminate between earthquakes and quarry blasts, Bull. Seism. Soc. Am., 80, pp. 2143--2160, 1990.
\\

Kanasewich, E.R.: Time Sequence Analysis in Geophysics, The University of Alberta Press, Edmonton, Alberta, Canada, 1973.
\\

Oppenheim,A.V.: Generalized Linear Filtering, in B. Gold and C.M. Rader: ``Digital Processing of Signals'', Chap. 8, pp. 233--262, McGraw-Hill, Inc., U.S.A., 1969.
\\

Oppenheim, A.V., R.W. Schafer: Digital Signal Processing, Chap. 10, pp. 480--531, Prentice-Hall, Inc., EnglewoodCliffs, New Jersey, U.S.A., 1975.
\\

Press, W.H., S.H. Teukolsky, W.T. Vetterling, B.P. Flannery: Numerical Recipes in FORTRAN: the art of scientific computing, Cambridge University Press, U.S.A.,1992.
\\

Rabinowitz, N., M. Joswig: Detection, localization and identification of local seismic events by a small array in Israel, Scientific progress report, submitted to the German-Israel Foundation for Scientific Research \& Development, Israel/Germany, 1995.
\\

Smith, A. T.: Discrimination of explosions from simultaneous mining blasts, Bull. Seism. Soc. Am., 83, pp. 160--179, 1993.
\\

Stump, B.W.: Nuclear Explosion Seismology: Verification, Source Theory, Wave Propagation and Politics, U.S. National Report to International Union of Geodesy and Geophysics 1987--1990, Reviews of Geophysics, supplement, pp. 734--741, 1991.
\\

[15] W\"{u}ster, J.: Discrimination of chemical explosions and earthquakes in central Europe --- a case study, Bull. Seism. Soc. Am., 83, pp. 1184--1212, 1993.
\\

[16] W\"{u}ster, J.: Diskrimination von Erdbeben und Sprengungen im Vogtland und Nordwest-B\"{o}hmen, Berichte des Instituts f\"{u}r Geophysik der Ruhr-Universit\"{a}t Bochum, Reihe A, Nr. 42, Civillier Verlag, G\"{o}ttingen, Germany, 1996.

%
%
%
%

\clearpage

\chapter*{9 Appendix}
Table 4: List of events used.\\
\includegraphics[width=1.0\textwidth]{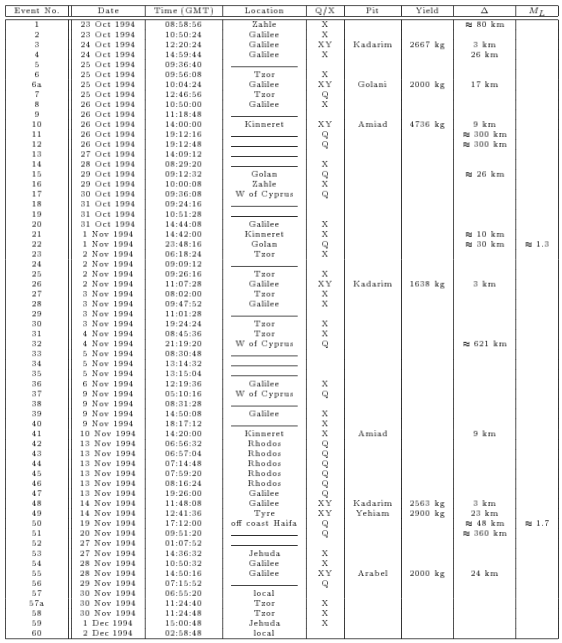}
\clearpage
\includegraphics[width=1.0\textwidth]{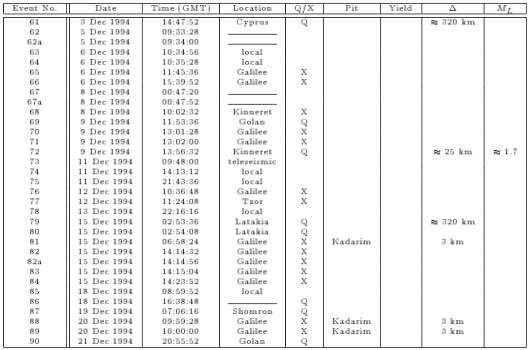}

\clearpage

\begin{figure}
\begin{center}
\includegraphics[width=1.0\textwidth]{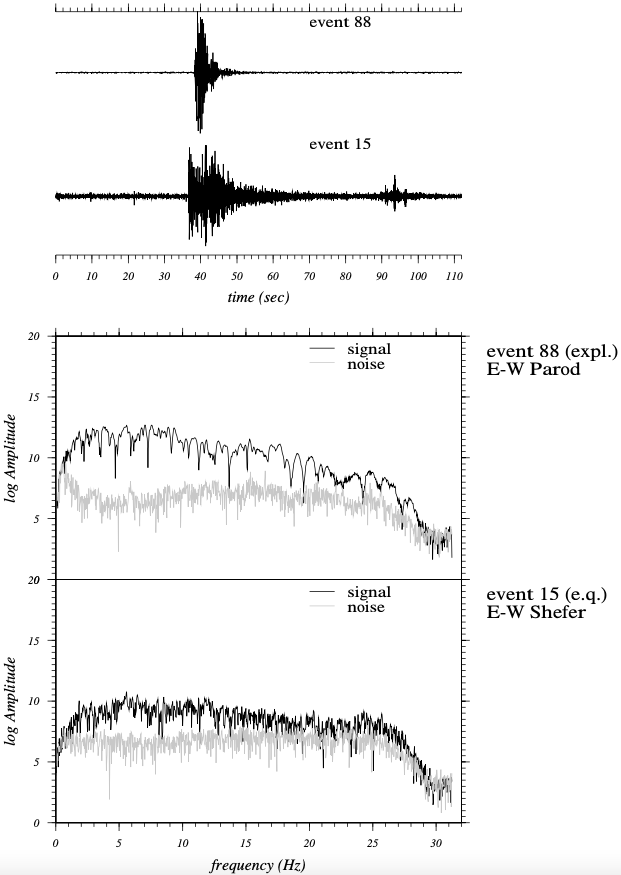}
\caption{Examples of an open pit blast and an earthquake respectively}
\end{center}
\end{figure}

\clearpage

\begin{figure}
\begin{center}
\includegraphics[width=1.0\textwidth]{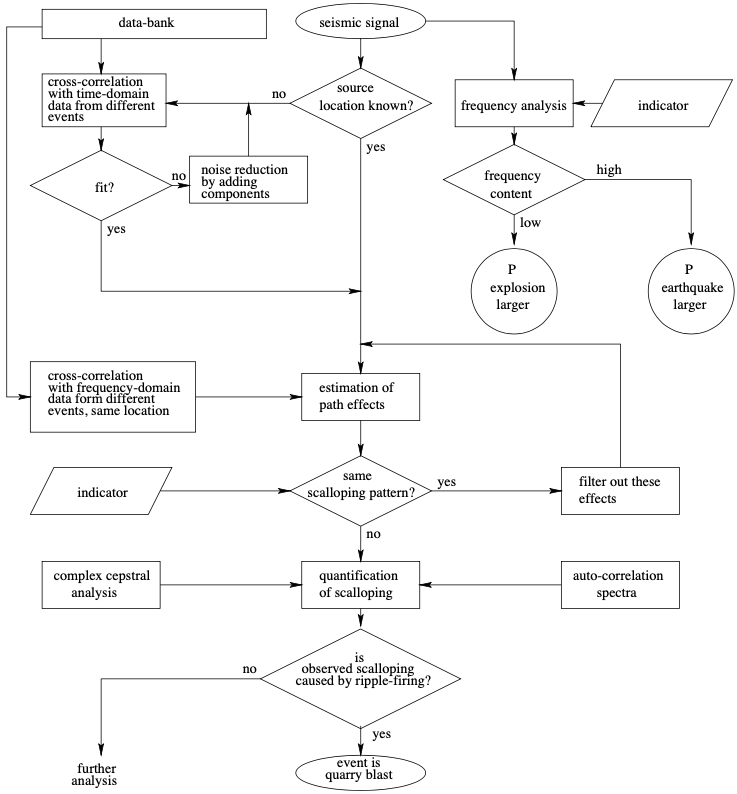}
\caption{Flow-chart of the analysis system}
\end{center}
\end{figure}

\end{document}